# Large Language Models for Education: ChemTAsk – An Open-Source Paradigm for Automated Q&A in the Graduate Classroom

Ryann M. Perez,[a]* Marie Shimogawa[a], Yanan Chang[a], Hoang Anh T. Phan[a], Jason G. Marmorstein[a], Evan S. K. Yanagawa[a], and E. James Petersson[a]*

* Indicates Corresponding Author

[a] Department of Chemistry, University of Pennsylvania, Philadelphia, Pennsylvania 19104, USA

**Email:** ryper@sas.upenn.edu, ejpetersson@sas.upenn.edu

**Abstract**

Large language models (LLMs) show promise for aiding graduate level education, but are limited by their training data and potential confabulations. We developed ChemTAsk, an open-source pipeline that combines LLMs with retrieval-augmented generation (RAG) to provide accurate, context-specific assistance. ChemTAsk utilizes course materials, including lecture transcripts and primary publications, to generate accurate responses to student queries. Over nine weeks in an advanced biological chemistry course at the University of Pennsylvania, students could opt in to use ChemTAsk for assistance in any assignment or to understand class material. Comparative analysis showed ChemTAsk performed on par with human teaching assistants (TAs) in understanding student queries and providing accurate information, particularly excelling in creative problem-solving tasks. In contrast, TAs were more precise in their responses and tailored their assistance to the specifics of the class. Student feedback indicated that ChemTAsk was perceived as correct, helpful, and faster than TAs. Open-source and proprietary models from Meta and OpenAI respectively were tested on an original biological chemistry benchmark for future iterations of ChemTAsk. It was found that OpenAI models were more tolerant to deviations in the input prompt and excelled in self-assessment to safeguard for potential confabulations. Taken together, ChemTAsk demonstrates the potential of integrating LLMs with RAG to enhance educational support, offering a scalable tool for students and educators.

**Main Text**

**Introduction**

Large language models (LLMs) have advanced considerably in recent years, displaying remarkable performance across various scientific tasks. OpenAI's recent releases of ChatGPT have demonstrated outstanding results on benchmarks and standardized tests, such as the GRE, the SAT, and the bar exam (1). Building on these advancements, open-source models such as Llama 3.0 from Meta are quickly approaching performance benchmarks of frontier models despite having fewer parameters (2). These achievements make LLMs appealing tools for scientific discovery and education. For example, within our research group, generative artificial intelligence (AI) and LLMs have enabled protein stability prediction and biologic aggregation prediction (3, 4). Within the broader field of biological chemistry, LLMs have enabled development of new pharmaceuticals through predictive modeling, molecular synthesis, and experimental planning (5-7). In the education domain, LLMs have proven useful for tasks such as simplifying teaching material preparation, educational planning, and creative brainstorming (8-10). The success of LLMs in these tasks underscores their potential for utilization in graduate level educational settings (11).

Recent investigations into LLM practical uses in education show that they are capable of problem-solving in the natural sciences and engineering, sometimes outperforming humans, suggesting significant potential for their integration into advanced educational settings. For example, when ChatGPT 3.5 was posed with undergraduate- to graduate-level engineering and natural science questions and graded by human reviewers, Balhorn *et al.* noted that the LLM was able to answer questions somewhat proficiently, but with performance deteriorating as the questions increased in difficulty (10). In chemical engineering, ChatGPT 3.5 was utilized to aid students in the design of distillation columns (12), and as a tool to decompose complex problems (13). In the field of natural language processing, ChatGPT outperformed human researchers in idea generation when evaluated by human judges (9). ChatGPT has also demonstrated that it can outperform humans on text-annotation at a fraction of the cost (14). As a coding educator, ChatGPT based systems have shown efficacy in aiding programming proficiency, especially for students with poor performance (15, 16). Within the field of chemistry, ChatGPT 3.5 demonstrated potential as a brainstorming tool for creating greener chemistry procedures and as a tool to practice troubleshooting skills (17, 18). In a more recent study, ChatGPT-4 was shown to be able to pass non-project questions for many science and engineering disciplines in an undergraduate curriculum (19). The empirical and observational evidence from these studies suggests that LLMs are broadly useful for answering questions and problem solving, but LLM usage in an advanced, interdisciplinary classroom comes with potentially critical issues that need to be addressed.

LLMs have limitations that have potentially prevented their wide adoption. Due to the significant time required for LLM training, their training data can become outdated over time, posing problems in classrooms covering recently published literature. Another inherent limitation of LLMs is their tendency to produce "hallucinations" — information that an AI confabulates — and falsifications (20). For ChatGPT, a common complaint is that hallucinations generate links to nonexistent journal articles (21). Falsifications can also arise when training data is intentionally manipulated with incorrect facts in a process called poisoning (22). Further, when a model has limited training data on a certain topic, the model is prone to creating arbitrary information when prompted multiple times with similar prompts (23, 24). Training data or processes can be biased as well which can be subtly reflected in the way these models interact with the user (25). Due to these issues, the use of LLMs as tools within education may falter, as students may receive incorrect, misleading or harmful information detrimental to their learning success. One potential solution is to supply LLMs with additional information relevant to the topic of interest, a process called retrieval-augmented generation (RAG). RAG offers a rapid and inexpensive solution to these problems as the model of interest does not need to be retrained (26-28). By adding supplemental context to a query, an LLM can more appropriately respond with accurately sourced information (28). This technique can decrease usage expenses as well. For example, RAG has proved useful in bolstering the answering capabilities for Llama 2 models with 7B and 13B parameters which are significantly cheaper to run than their proprietary counterparts (29).

We aimed to study LLMs augmented with RAG in the context of an advanced biological chemistry classroom. We define biological chemistry as the study of living molecular systems using chemical-scale approaches. Over a period of nine weeks in the 2024 spring semester offering of Biological Chemistry II (**Chem5520**) at the University of Pennsylvania, we investigated the feasibility of using LLM systems for student question and answer (**Q&A**) sessions. Chem5520 covered topics ranging from generating synthetic DNA mimetics to tissue and live organism imaging. Course work consisted of short weekly quizzes, project proposals grounded in primary literature, and an oral final exam. These assignments asked for student creativity and blended a student's fundamental understanding with modern techniques to design novel experiments.

Given the potential for complex queries from students and the inherent limitations of LLMs, we developed the Chemistry Teaching Assistant system for knowledge (**ChemTAsk**). ChemTAsk is an open-source pipeline that facilitates LLM-student communication through an educator monitored system. This technology leverages the capabilities of LLMs to provide answers to precise and potentially creative questions grounded in academic literature and class materials while safeguarding against hallucinations and other LLM hazards. After the semester, we evaluated ChemTAsk and teaching assistant (TA) responses to the same set of representative queries. We found that ChemTAsk is proficient at answering queries involving creativity while TAs gave responses more tailored to class material. After our initial pilot study, we aimed to understand how to best implement simpler LLMs to lower the potential cost for educators by investigating a combination of RAG techniques and prompt engineering schemes. A common and relatively inexpensive way to evaluate LLMs is through a multiple-choice test (30, 31). For example, the Google-Proof Question and Answer benchmark (GPQA) demonstrates LLMs' ability to solve novel problems in natural science. The GPQA benchmark includes multiple-choice questions that are not present in model training data. While the GPQA is difficult for non-domain experts, frontier LLMs have shown some proficiency in solving them (32). Taking inspiration from the GPQA benchmark, we created a novel 40-question biological chemistry benchmark to assess the reasoning performance of proprietary and open-source models. We additionally explored ways to potentially safeguard against misinformation in model responses. Through these experiments, we found that proprietary models outperformed open-source models in accuracy and hallucination prevention at the cost of monetary resources. Taken together, our findings suggest that ChemTAsk is a potentially transformative tool for higher education.

**Results and Discussion**

**ChemTAsk Usage and Observations**

The study's design is depicted in **Figure 1a**. Following student recruitment and consent, the pilot study was initiated over a 9-week period. The instructor (Prof. Petersson) and the TAs assigned to the class (Shimogawa and Chang) were blinded to the students' use of ChemTAsk so as to avoid any influence on course interactions or grading. Only the study administrator (Perez) was able to observe and provide support for ChemTAsk usage. The goal of the initial study was to garner data on real world student and ChemTAsk interactions, assess these interactions using human reviewers, and provide a blinded evaluation of ChemTAsk's performance. The overall mechanism of ChemTAsk is explained in the methods section and in **Figure 1b**. A student can email the ChemTAsk server and would receive an answer accompanied by a reference document that matched their query. Given the anticipated capabilities of ChemTAsk, several assignments were restructured from previous offerings of Chem5520 to ensure that students could not simply rely on the LLM to provide answers, most notably the weekly quizzes and final oral exam. The weekly quizzes featured questions requiring application of class material to the students' own research projects. Finally, the oral exam allowed students to use ChemTAsk or any other tool of their choosing in preparing notes, but the final required students to respond in real time to the examiner's questions. These oral exam questions sampled from the semester's material. Some of these materials are referenced in the student-ChemTAsk interactions.

During the duration of the study, 233 student-ChemTAsk interactions were recorded (**Dataset S1**). Student queries frequently requested topics discussed in class and hypotheticals for creating final projects and proposals. Word clouds and word frequency were generated for both the body of the email and response from ChemTAsk (**Figure S1 and S2**). Most student emails contained queries regarding proteins and amino acid sequences, topics largely covered in the class, as well as the specific protein LexA which can be found within the final exam. The word frequency from ChemTAsk mirrored that of the queries from students, with the exception of words related to citing a source which is a direct consequence of using the Assistants functionality from OpenAI's application program interface. The average response time for ChemTAsk was 2 minutes from the initial email to the student receiving feedback. ChemTAsk was utilized 24.7% of the time on the day before the due date of online quizzes (**Figure S4**) indicating it was likely used for these assignments. In fact, questions from quizzes and the final exam were often found in user queries. The top five users (of 19 total) comprised 67.7% of the engagement with ChemTAsk (**Figure S5**) indicating that these students may have found the service particularly useful after initial interactions. The top five users, and the cumulative usage of the other 14 users, remained relatively consistent over time; however, usage increased particularly as the final exam approached (**Figure S6**). While the study administrator observed interactions over the duration of the class, no obvious errors in factual information were encountered, however, it was observed that the system did occasionally produce answers which were unrelated to the query.

**Evaluating Human and ChemTAsk Performance**

After the completion of the semester, we asked whether ChemTAsk performed as well as a human TA on a representative Q&A dataset. We assigned five current and former graduate students with TA experience and a background in biological chemistry (Shimogawa, Chang, Phan, Marmorstein, and Yanagawa) to write responses to a set of anonymized representative queries posed to ChemTAsk by the Chem5520 students. TA responses and ChemTAsk responses were then graded by a set of graduate students with experience in biological chemistry using a set of five criteria. The results of comparing human answers to ChemTAsk answers are detailed in **Figure 2a**. ChemTAsk performed similarly on average on the set of 50 queries, but there were several notable differences between ChemTAsk and human TAs. The top two scoring query and answer pairs for TA responses versus ChemTAsk responses are detailed in **Table 1**.

ChemTAsk could produce creative and correct answers for biological chemistry content as exemplified by Q31, which was embedded within question 2 on the final exam. This question discusses the SOS response in bacteria which is a mechanism responsible for antibiotic escape (33). Theoretically, through gene editing with technologies such as CRISPR and fluorescence microscopy, one could monitor how changing genes within the SOS response affects cellular machinery. ChemTAsk replied with an outline for this technique (**Table 1**). As judged by our panel, ChemTAsk gave a good answer and received the best score from our reviewers. However, comparison to the TA response was hindered by the fact that the TA could not reasonably provide an answer in the time frame provided. The lengthy response required for this final exam question was beyond their expertise in the time allotted. Indeed, ChatGPT itself has demonstrated that it can be a helpful brainstorming tool, and the ChemTAsk response was an outline that could be further elaborated by the student (34). Another instance in which ChemTAsk excelled was in its description of the 'bump-and-hole' strategy in Q13. The bump-and-hole strategy is a chemical genetics approach which modifies a protein to create a 'hole' while modifying its interaction partner, a small molecule ligand, to create a complementary 'bump.' This technique can be used to probe gene function and protein pathways (35). ChemTAsk gave a well-scored explanation, but the TA who answered this particular question did not perform as well. Q13 also highlights a strength of LLMs: they have shown superior knowledge retrieval in interdisciplinary fields whereas a human may have to re-learn topic material to properly respond (36). The TAs had their own strengths, as sometimes ChemTAsk would not recognize the query. Particularly in Q10, which discussed extracellular sensors, ChemTAsk received an unhelpful resource which resulted in the system explaining unrelated concepts. The TA responded appropriately and therefore was more favorably reviewed by experts. In Q16, a query exploring fluorescent probes discussed in class, the TA had more relevant knowledge and could answer this query in more specific detail. ChemTAsk seemed to struggle since the question was about particular fluorophores not discussed in class. In general, the ChemTAsk system was better at answering questions that were creative and required multiple domains of knowledge, while the TAs were more precise in their responses, a finding which aligns with previously published observations on AI versus human experts (9, 34, 37). This finding illustrates one of the challenges in providing TA support for a class like Chem5520, where graduate students may still be refining their own knowledge in a specific domain while teaching class material. Historically, the instructor (Petersson) has managed the Chem5520 Q&A online discussion boards, thereby providing a broad scope of knowledge demonstrated by ChemTAsk and the precision of the human TA. However, instructor availability is generally less than TA availability and clearly less than LLM availability.

Further analysis uncovered opportunities to enhance both ChemTAsk's performance and our ability to critically assess it. **Figure 2b** illustrates performance at the individual question level. Notably, on two out of five metrics, ChemTAsk performed significantly better than the TAs on average, specifically in understanding intent and overall correctness (ChemTAsk scored 4.5 out of 5 vs TAs' 4.2 for question 1 and ChemTAsk scored 4.7 vs TAs' 4.5 for question 5, $p < 0.05$, t-test). This is surprising, as ChemTAsk performed poorly on Q10 due to misunderstanding the intent of the question. This issue could be potentially mitigated with a reworded query so that the document retrieval system could pull a more relevant source. Thus, some coaching on prompt design and scrutiny of ChemTAsk responses could improve its performance for end users (38). Although judges were blinded to the source of responses (ChemTAsk or a TA), ChemTAsk exhibits a recognizable response style likely due to its use of ChatGPT-4-turbo as the core model. ChatGPT-4-turbo extensively uses Markdown in the set of responses explored in this study, making it stylistically distinct from our TAs and humans in general (39). We investigated whether bias was statistically evident in the evaluation of the model. To assess inter-evaluator reliability, Krippendorff's alpha was calculated for each question averaged over the entire dataset (**Figure 2c**). When responses were written by TA's, the judges agreed on the score given. However, for ChemTAsk, scores were more mixed, and judges agreed less often. Human bias, particularly beauty and authority bias, is well-documented in such evaluations (40). That is, the text generated from ChemTAsk may look more organized and authentic to a human annotator than TA-generated content which is why ChemTAsk may achieve statistically better scores. Alternatively, some evaluators may exhibit skepticism and distrust towards AI when grading responses. A few studies found that 9-10% of higher education educators on Twitter/X exhibited fears or concern for the technology (41, 42). Therefore, an anti-AI bias is plausible within our judging panel. Based on the evidence discussed, evaluator bias likely plays a role in the scores given to ChemTAsk and future studies could specify the output tone to mitigate these effects, especially for the purposes of post-study evaluation.

**Student Perceptions**

Student experiences with ChemTAsk were generally positive. Of the 17 students who participated in the study and completed the exit survey, 14 reported utilizing ChemTAsk. Respondents who used it generally thought the answers were correct, a sentiment echoed by other studies and our observations of ChemTAsk-student interactions (10, 20). However, since our specific evaluations of responses from ChemTAsk identified errors, this suggests that students were not always able to identify incorrect responses, one of the chief concerns about implementing AI tools in the classroom. For this reason, the ChemTAsk pipeline featured oversight of the answers by the study administrator (Perez) and in future implementations where blinding is not necessary for study purposes, oversight could be provided by the instructor and TAs. Yet, we did not observe incorrect answers beyond obvious errors in interpreting the query suggesting that incorrect details may be subtle. When students were asked if the system provided relevant sources, the students viewed the sources as correct given the context of their query. ChemTAsk also had other clear advantages. Firstly, ChemTAsk was available at all hours, unlike a human TA. Additionally, owing to the two-minute average response time, students thought ChemTAsk answered faster than a TA which would potentially allow for students to ask questions that they thought were questions not worth the TA's attention. Given these sentiments and Q&A analysis, it is clear that LLMs have a role in advanced classes. Further, students thought ChemTAsk responses contained satisfactory detail, that it was easy to use, and that it enhanced their understanding of biochemistry (**Figure S3**). While the speed of ChemTAsk is slower than ChatGPT due to the RAG process, some students saw it as a reasonable alternative to a premium subscription to ChatGPT Plus. A subscription would cost $20 USD per student as of the May 2024, the time of this study.

The students' short responses echo the survey results. When students were asked what they liked about ChemTAsk, one student reported that "it [was] much easier to ask questions as compared to getting help from a real TA," and another said "I liked that the resources it could pull from were directly the ones used in class. Thus, it would only give us information directly relevant to what we were learning." These statements underline the potential for customized learning in the graduate classroom. While similar RAG systems such as Perplexity AI draw from a corpus of data on the internet, the corpus generated by educators for their classroom is smaller and may enable more relevant knowledge retrieval from the class. Criticisms often stemmed from how the system was built, for example: "When I mistyped [ChemTAsk] in the subject line, I didn't receive a response," or the fact that there was no chat option. When asked about their opinions on generative AI in the classroom, some students thought that generative AI has a "huge potential in education" and could be "used to enhance learning." One student said "I thought that it was a great supplement to reaching out to the professor or the TAs! It was helpful not feeling like I am annoying the bot." While the human-to-human exchange is limited due to the availability of the TA or professor, the human-to-AI interaction is limited by nominal computational and monetary costs. Student testaments and responses to questions are detailed in the supplementary information.

**Instructor Perceptions**

Throughout the course, instructors observed that students increasingly utilized ChemTAsk to respond to quiz questions and class assignments. Upon grading these submissions, it became apparent that certain answers exhibited a distinct style suggestive of AI-generated content, characterized by more formal and structured language compared to the more colloquial tone typically found in student responses (39). ChemTAsk provided well-articulated answers and performed consistently well on questions focused on conceptual understanding, delivering correct responses with high accuracy. However, there was a trend of students submitting AI-generated content verbatim into their work. They also observed that grading became harder as responses increased in length after the implementation of ChemTAsk. This prompted educators to scrutinize such submissions more closely to identify inaccuracies or misconceptions, as the responses from ChemTAsk, while generally insightful, sometimes contained significant errors.

**Improving ChemTAsk**

Given the initial success of the ChemTAsk system, we aimed to examine how the core LLM of the system can be improved or reduced in cost. Over the course of the semester, $58.76 USD in API tokens were used for a system with 32 users (**Figure S6**). For the 233 Q&A pairs, this equates to a cost of $0.25 per interaction. However, in a larger class with more frequent usage, the total cost could become burdensome as students familiarize themselves with LLMs. Therefore, we hypothesized that we could utilize a smaller model augmented with RAG, such as the open-source Llama 3.0 8B parameter model (**Llama 8B**), which fits on a single consumer GPU, and ChatGPT4o-mini, which is one of the best commercial models with lower costs compared to ChatGPT-4o or ChatGPT-4-turbo. Rather than use repeated cycles of short answer questions and human expert evaluation to assess model performance, we developed a 40-question multiple choice test that covered most topics presented in class and current research (Supplementary File). Since the exam was an original work created after the training cutoff of these models, the answers to these questions should be absent from the training data and we should be assessing model reasoning capabilities and core knowledge. The accuracy of ChemTAsk on this set of questions (using ChatGPT-4-turbo given whole documents via their Assistants functionality) was 67.5% (**Figure 3**). For comparison, the best model from Perplexity AI achieved a mean score of 65.8% averaged over three rounds as deterministic model outputs were not possible. This demonstrates that ChemTAsk can perform on par or better with relatively fewer resources as ChemTAsk does not have access to the internet.

We next examined if more cost-effective LLMs could reach the level of performance of the largest LLMs. It has been shown that using different prompts can improve the performance of LLMs (38, 43). Therefore, we asked ChatGPT-4o to create several prompts based on our initial input (see Methods). We then evaluated several combinations of prompts and a number of 500-word context sections given to the model. While more information could aid the model in determining the correct answer, their context windows are limited, and it has been shown previously that longer context lengths may lead a LLM to lose key information (44). Strikingly, the initial prompt designed by the authors performed the worst when given any context via RAG for Llama 8B (**Figure 3a**, Prompt 1, 45% for 1 context section). The six prompts generated by ChatGPT-4o performed better with 65% being the most likely outcome, and two sections (1000 words) performed the best for all prompts with the exception of prompts 5 and 1. Comparing prompt 1 with the other prompts indicates that simplicity may be desirable for accuracy. Prompts 5 and 1 also do not explicitly mention listing a letter which could lead to uncertainty in the LLM output. For ChatGPT-4o-mini (**Figure 3b**), the model was much more tolerant to perturbations in the input prompt with similar accuracy to Llama 8B. This volatility as a function of the variation in input prompts of Llama 8B may indicate that the model is unsure about its responses, and this feature may be an inherent limitation of smaller models (23).

We found that monitoring the emails of student-ChemTAsk interactions for hallucinations or incorrect information was challenging and time-consuming. Recent studies have shown that LLMs have the ability to self-assess their own outputs for confabulations, or fabrications that appear plausible. This methodology could be used to potentially limit false or incorrect information that reaches the student. To test model self-assessment, we implemented the probability true (**P(True)**) metric from Farquhar *et al.* (23) and adapted it for OpenAI models (see Methods). This metric asks the LLM to brainstorm answers at a higher temperature, or creativity level, and assess whether the brainstorm answers match the low temperature or most likely answer. **Figure 3c** displays that for the set of Llama models, both performed poorly compared to OpenAI models on self-assessment on the training set. This trend holds true for a holdout set as well (**Figure 3d**). While the Llama 3.0 family of models are the most adaptable as they can be more easily customized and adapted for a course, the OpenAI family of models are more consistent and have the ability to self-assess in the realm of biological chemistry.

For future implementations of ChemTAsk, it is important that the self-assessment metric used is compatible with processing paragraph-length responses. In addition to P(True), semantic entropy can be utilized in this scenario which has been shown to outperform P(True) on detecting confabulations at the paragraph level. Semantic entropy measures the relatedness of multiple responses from a LLM to determine if individual factoids are confabulations (23). That is, semantically distant responses from similar questions may indicate that the LLM may be producing a confabulation. We prompted ChatGPT-4o to produce short responses because not only did students give the amount of detail generated by ChemTAsk a less than perfectly favorable rating, but also because the implementation of semantic entropy can become lengthy in time and costs as the length of the output text grows. **Figure S7** displays semantic entropy and P(True) distributions for these models on paragraph length responses. For P(True), ChatGPT-4o was certain about the vast majority of responses. However, semantic entropy revealed a wider distribution of uncertainties. Upon closer inspection of the most uncertain responses judged by P(True) and semantic entropy, we found that they belonged to two categories: potential confabulations or factoids with multiple valid answers. For instance, if the model responds differently to the same question, it may indicate uncertainty. However, in cases where multiple answers are correct, the model's varied responses may all still be acceptable, but the uncertainty metric remains high. In future implementations of ChemTAsk, this metric can be provided to students with additional context to help interpret uncertainty. By informing students when a response has high uncertainty, they can ask follow-up questions to clarify ambiguities, explore alternative answers, or identify potential confabulations.

Based on the results from our experiments with proprietary and open-source LLMs, OpenAI models are desirable for use as the core LLM for its tolerance to prompt changes and its ability to detect confabulations. As of November 2024, utilization of ChatGPT-4o-mini would correspond to 10-20 times cost reduction according to OpenAI's current model pricing for ChatGPT-4-turbo and ChatGPT-4o-mini. These results demonstrate that RAG in conjunction with prompt engineering can improve a smaller model's accuracy, thereby potentially reducing the overall cost to educators.

**Conclusion**

In this study, we showed that an LLM-based system such as ChemTAsk is potentially beneficial to students in the graduate-level science classroom for creative problem solving. ChemTAsk responses were perceived as being generally correct by students, and our panel of judges agreed with this sentiment. The 24-hour availability of ChemTAsk and its ability to respond to repeated questions without the student being concerned for unfairly burdening the TA are clearly valuable. However, this initial study also identified areas for improvement. Although ChemTAsk excelled in Q&A tasks, there were clearly instances where the system did not recognize the intent of a query and instances where the model was subtly incorrect as revealed by evaluators. We also demonstrated through a novel biological chemistry benchmark that the core LLM, ChatGPT-4-turbo, could be replaced with cost effective versions such as ChatGPT-4o-mini or ChatGPT-4o. Additionally, seeing the need for additional safeguards for student education, we implemented P(True) and semantic entropy to detect potential confabulations. For broader use of LLMs in education, we have provided the original implementation of ChemTAsk is available on our GitHUB: https://github.com/ejp-lab/EJPLab_Computational_Projects/tree/master/ChemTAsk. We will continue to improve ChemTAsk in future course offerings and look forward to its adoption and modification by other educators.

**Limitations**

While this study focused on text Q&A, biological chemistry contains many pictographic elements (e.g. molecular structures, protein visualization, reaction mechanisms, etc.). Therefore, students could not insert images from biomolecule visualization software (e.g. PyMOL(45)) to help the model to understand certain problems. Current frontier models such as ChatGPT-4o can utilize vision capabilities and will be the focus of future studies (1).

**Materials and Methods**

**Student Recruitment**

Students of the class of CHEM5520 were notified of their eligibility to participate in the study at the beginning of the semester. Recruitment began after approval of the study through the University of Pennsylvania's internal review board (IRB Protocol Number: 855207). Interested students could opt-in to using the service, having their responses recorded, and/or having their grades recorded. Students who participated in the study could use ChemTAsk over the duration of the remainder of the course (9 weeks). Blank recruitment documents can be found on our GitHub: https://github.com/ejp-lab/EJPLab_Computational_Projects/tree/master/ChemTAsk

**Document Storage**

Publications and audio transcripts were collected on a weekly basis from the course instructor. Each document was split into 500-word chunks and then embedded using the text-embedding-3-large model from OpenAI's API. Embeddings were then stored for later use and were automatically updated as the class publication list was updated.

**ChemTAsk System**

The general workflow is depicted in **Figure 1b.** Briefly, the student consent forms were used to generate spreadsheets containing their first and last names for the privacy filter, and a preferences spreadsheet to dictate if their responses were recorded. Students could email the CHEM5520 email with their query at any time after receipt of the consent form. The Gmail and Pub/Sub APIs were used to notify the server of incoming emails containing the phrase "chatgpt" in the subject line. Student emails containing personally identifying information were identified using a strict privacy filter and replaced with the string "<FILTERED>". The student query was then embedded using the text-embedding-3-large model from OpenAI's API and compared to document embeddings using cosine similarity (**Figure 1b**). The best matching document text was then retrieved and supplied to OpenAI's Assistant API to answer the student query. The response from the assistant was routed back to the student along with the resource it used to answer the question. For the duration of the study, ChatGPT-4-turbo was used as the generative model. Interactions between students and ChemTAsk were monitored to minimize potential inaccuracies. Scripts for running the ChemTAsk server and data collection can be found at https://github.com/ejp-lab/EJPLab_Computational_Projects/tree/master/ChemTAsk

With the exception of Ryann Perez, all educators (Marie Shimogawa, Yanan Chang, E. James Petersson) in the study were blinded to student interactions with ChemTAsk until all grades were submitted. Educators were asked to reflect on their experience with grading assignments that were potentially aided by the ChemTAsk system. Students were able to use ChemTAsk in all parts of the class with the exception of during the final oral exam, however, they were allowed to see the final exam questions beforehand and utilize ChemTAsk to prepare.

**Expert TA Recruitment for Response Generation**

Five expert TAs were recruited for response generation. Expert TAs are defined as those that had previously served as a TA and taken Chem5520 or its equivalent. Each of the expert TAs were asked to answer 10 questions derived from a random stratified split of questions from each of the 9 weeks of the course. Expert TAs were asked to respond to each query and provide a reference document. 30 minutes were given for each query to give TAs ample time to respond and provide a reference document. The Q&A responses are recorded in supplementary information.

**Expert in Chemical Biology Recruitment and Grading**

Experts in chemical biology are defined as those who have previously passed a graduate level biological chemistry class. Recruitment took place over email with a nominal $2.50 USD per question advertised. 14 experts were recruited in total. Experts were asked to answer the following questions for each query on a 5-point Likert scale (adapted from ref.(46)):

1. The response recognized the query's intent and answered appropriately.
2. The response provided a document/citation/link to a resource that would help answer the query.
3. The response provided relevant information that you would have included given the query.
4. The response did not contain too little or too much detail.
5. The response contained generally correct information.

Query/response pairs were randomly distributed, the response originator (i.e. ChemTAsk or the human TA) was hidden, and each response was graded three times by separate reviewers. For each question, Krippendorf's alpha analysis was performed to determine overall agreement.

**Generation of Biological Chemistry Q&A Dataset and Evaluation of ChemTAsk**

Forty multiple choice questions were written from existing knowledge of literature. Questions and answers were then edited and evaluated by the coauthors of this study. Models were given zero to three sections of context with each section consisting of 500 words from the most relevant section of text. The following prompt was given following the multiple-choice question and context:

"Answer the multiple-choice question provided above. PROVIDE ONLY A SINGLE CAPITAL LETTER AFTER "Answer". The information provided within [CONTEXT] may or may not help you answer the question."

The model output was then stripped of whitespace. If the answer included any extra characters, the intended answer was extracted manually. The multiple-choice benchmark can be found at https://github.com/ejp-lab/EJPLab_Computational_Projects/tree/master/ChemTAsk

**Inference with Llama Models on Multiple Choice Test**

Code for running Llama3 8.0 model can be found on our GitHub at https://github.com/ejp-lab/EJPLab_Computational_Projects/tree/master/ChemTAsk. Inference for Llama3-instruct-8B was run on a single NVIDIA A6000 GPU. Temperature was set to 0 for all inferences.

**Inference with Perplexity AI Model on Multiple Choice Test**

Code for running Perplexity AI's model can be found on our GitHub at: https://github.com/ejp-lab/EJPLab_Computational_Projects/tree/master/ChemTAsk. The finetuned model 'llama-3.1-sonar-huge-128k-online' was used for all inferences with 'temperature' of 0. This inference script was run three times and averaged for a final score as the outputs were stochastic.

**RAG Implementation and Prompt Engineering for Smaller Models**

For the purposes of testing the smaller models (Llama 8B and ChatGPT-4o-mini), RAG was implemented by using the 500 word section that was most relevant to the query as the information provided to the model. The number of sections given varied from zero to the three most relevant sections (1500 words). Six additional prompts were tested generated by ChatGPT-4o:

**Prompt 1 (original)**: "Answer the multiple choice question provided above. PROVIDE ONLY A SINGLE CAPITAL LETTER AFTER \"Answer\". The information provided within [CONTEXT] may or may not help you answer the question."

**Prompt 2**: Please read the question and select the best answer from the options provided. Respond with only the capital letter (A, B, C, D, or E) that corresponds to your choice. Do not include any additional text.

**Prompt 3**: Answer the following multiple-choice question by selecting the option that best answers it. Provide your answer as a single capital letter. For example:\nQuestion: What is 2 + 2?\nA. 3\nB. 4\nC. 5\nAnswer: B.

**Prompt 4**: Carefully read the question and choose the most appropriate answer from the options. Provide only the capital letter corresponding to your choice.

**Prompt 5**: Think about the question and the context provided. After reasoning, provide your final answer as a single capital letter corresponding to the best choice. Do not include your reasoning in the answer.

**Prompt 6**: Use the provided context if it helps answer the question. If not, rely on your own knowledge. Provide your answer as a single capital letter (A, B, C, D, or E) without any extra text.",

**Prompt 7**: Select the correct answer from the options below. Ensure your answer is accurate. Provide only the capital letter corresponding to your choice.

Each prompt and RAG number combination was tested for the forty-question dataset.

**Assessment of LLMs on Multiple Choice by P(True)**

The multiple-choice questions were given to ChatGPT-4o to produce five slightly reworded questions with the same intended answer. Then, the multiple-choice questions were split into the train and test set randomly, however, all similar questions were kept together when stratified. That is, the test set questions should contain no reworded questions from the training set. This resulted in 147 questions in the train set and 46 questions in the test set.

Implementation of the P(True) uncertainty metric was implemented as described in ref. (23) with the exception of OpenAI models. As of this publication, the model output logits are only available for text generated by the model and cannot be inspected on the input tokens. Therefore, the model was allowed to answer 'A' or 'B' to the proposition that the proposed answer was true or false. For an answer of A, the output log probability was taken as is. For an answer of B, the complement of the probability of B was taken for P(True).

**Large Language Model Utilization**

ChatGPT-4o was used to help generate code for analysis, edit original drafts of the manuscript, and search for primary publications. In some instances, ChatGPT-o1-preview was used for complex code analysis.

**Acknowledgments**

This research was supported by a grant from the National Science Foundation to EJP (NSF CHE-2203909). MS thanks the Nakajima Foundation for scholarship funding. RP and EY were supported by the National Institutes of Health Chemistry Biology Interface Training Program (NIH T32-GM133398). JGM thanks the Vagelos Institute for Energy Science and Technology for fellowship support. The authors are grateful to the students of Chem5520 Spring 2024 for their participation in this study.

**Figures and Tables**

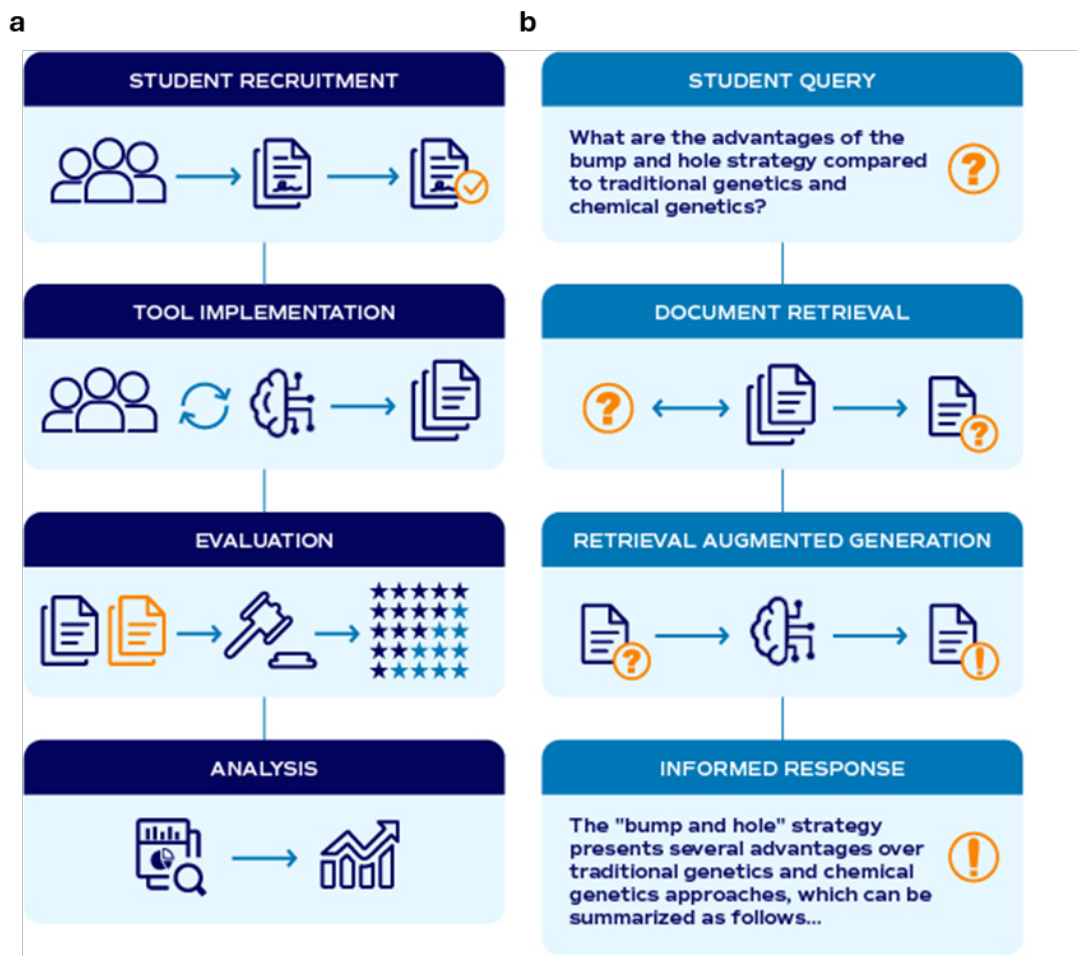

**Figure 1** – Workflows involved in this study **a)** Overall workflow for study execution. Students could interact with ChemTAsk and interactions were collected. Interactions were compared to human TA responses. The results of the study were analyzed, and improvements were explored. **b)** Overall workflow of ChemTAsk. Student queries are sent to a server which uses the query to match the best document found from class. The document and query are then sent to ChatGPT for RAG. This response is then sent back to the student with the document for further reading.

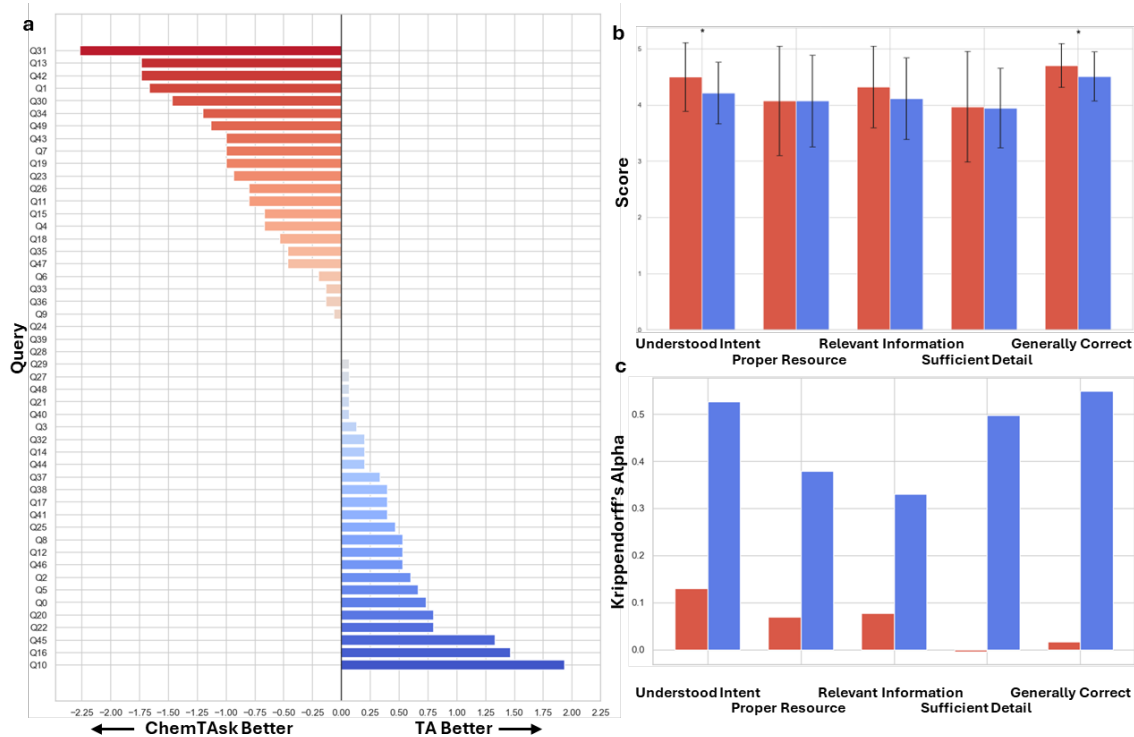

**Figure 2** Analysis of ChemTAsk's performance. **a)** ChemTAsk versus TAs on fifty representative queries. Negative numbers or red coloring indicate ChemTAsk performed better on a given query, blue indicates TAs performed better. **b)** Aggregate scores for ChemTAsk (red) and TAs (blue). Error bars indicate standard deviation. Astrix indicates a p < 0.05 for Wilcoxin Ranked U test. **c)** Krippendorff's alpha for ChemTAsk (red) and TAs (blue).

**Table 1.** Selected responses from ChemTAsk and TAs. Texts that were shortened for brevity are marked with [abridged]. A full list of unabridged responses is available in **Dataset S2**.

| Query # | Question | ChemTAsk Response | TA Response |
|---|---|---|---|
| Q31 | how can you Design <FILTERED> experiment that uses the PRIME editor to convert all the SOS promoters to dinG (fast) or sulA (slow) promoter sequences or to increase the LexA NTD-CTD linker length? | To design an experiment that leverages the PRIME editor to alter SOS promoters across dinG (fast) or sulA (slow) sequences, or to extend the LexA NTD-CTD linker length, the following steps can be envisioned based on the exam question outline:<br>1. **Experiment Overview**:<br>   - **Objective**: To observe the impact of modifying SOS promoter sequences to either fast (dinG) or slow (sulA) variants, or extending the LexA NTD-CTD linker length on the SOS response.<br>   - **Method**: Utilize the PRIME editor system, a CRISPR-based genome editing tool, for precise genomic alterations. [abridged] | This is beyond my expertise within the allotted time. |
| Q13 | What are the advantages of the bump and hole strategy compared to traditional genetics and chemical genetics? | The "bump and hole" strategy presents several advantages over traditional genetics and chemical genetics approaches, which can be summarized as follows:<br>1. **Merging Genetic Selectivity with Chemical Flexibility**: The bump and hole strategy combines the selectivity inherent in genetic techniques with the speed and precision of small molecules. This approach modifies proteins ("holes") and small molecules ("bumps") to fit together uniquely, allowing for highly specific interactions. This specificity is difficult to achieve with broader genetic or chemical methods that may affect multiple proteins or pathways simultaneously【5:1†source】.<br>2. **Speed and Dosage Control**: [abridged] | Traditional genetics start with random or site directed mutagenesis and a redundant biological pathway might cause the whole effort meaningless. Meanwhile, chemical genetics screen with chemical libraries for novel phenotypes. Compared to the traditional genetics, bump and hole strategy is a method in chemical genetics which allows for selective inhibition of a single engineered protein kinase without perturbing the other members of the family. |
| Q10 | How do the venus fly trap extracellular sensors work? | It appears there has been a misunderstanding. The document results do not mention the Venus flytrap's extracellular sensors but rather discuss technological advancements in measuring extracellular glutamate concentrations using genetically encoded FRET-based sensors. To clarify the operation of Venus flytrap extracellular sensors, I will provide an explanation based on existing biological knowledge…<br>[abridged] | In bacteria, there are different periplasmic proteins that can bind small molecule nutrients like this (e.g. glutamate, calcium, etc.), and they help to sequester these molecules and import them into bacteria. There are different sensor proteins for different small molecules, and there are sensing systems such as the extracellular calcium-sensing. An example of the extracellular sensor is the yellow fluorescent protein called Venus. This features a membrane anchoring domain and a single pass transmembrane domain. [abridged] |
| Q16 | Hi [ChemTAsk], Please explain the different types of fluorescent protein biosensors we discussed in class. When is it beneficial to use one compared to another? <REDACTED> | The document outlines three primary types of genetically encoded fluorescent protein biosensors used to monitor and measure protein activity or small molecule signals within cells. Here's a summary of each type mentioned…<br>[abridged] | In class, we discussed different types of fluorescent protein biosensors that involves fluorescent proteins such as green fluorescent protein and the likes (e.g. cyan and yellow fluorescent proteins). The first fluorescent protein biosensor strategy is "Ligand Induced Conformational Change" that often involves FRET… [abridged] |

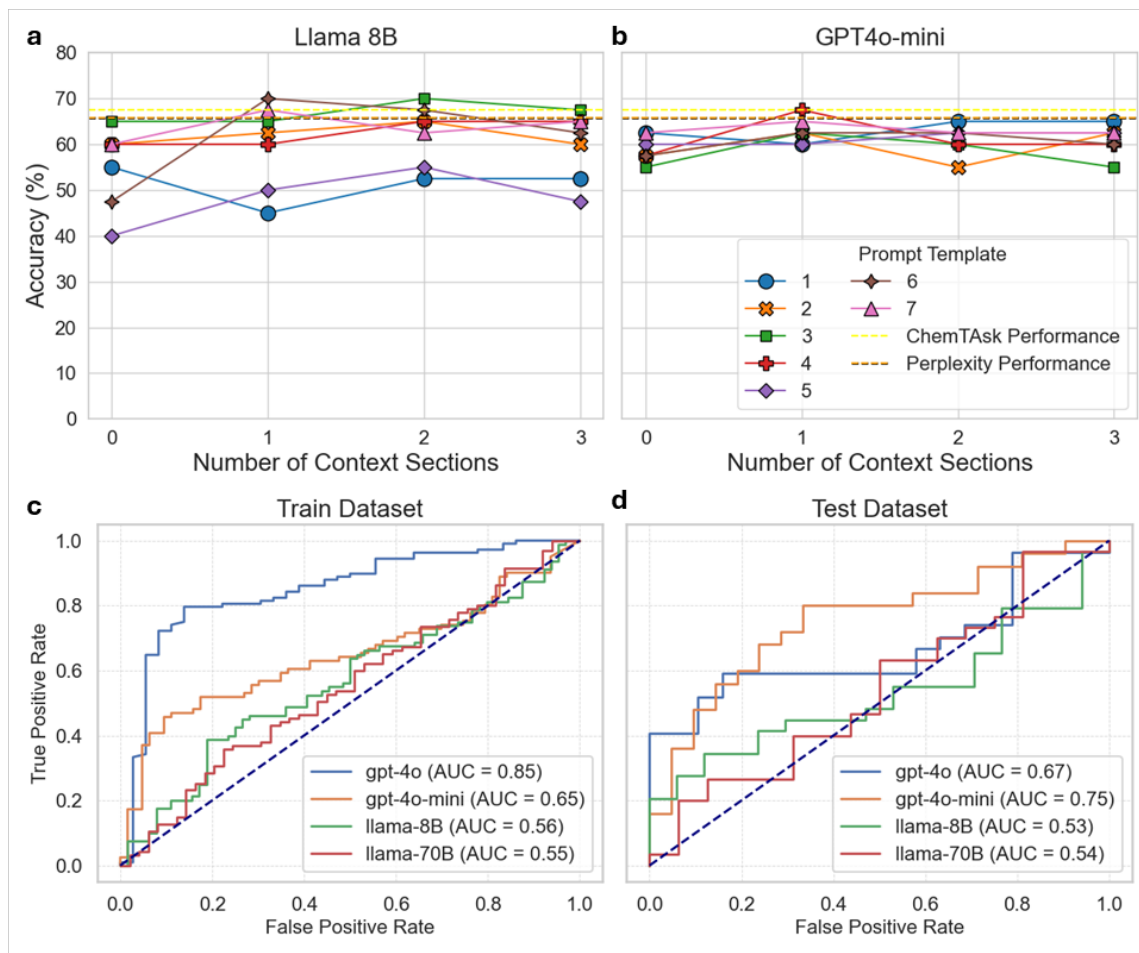

**Figure 3** – Assessing model performance on multiple choice tests and ability to self-asses. Model performance with number of context sections and prompt template given in **a)** for Llama 8B and **b)** for ChatGPT-4o-mini. Each combination of prompt template and number of chunks given to the model were evaluated for the 40-question biological chemistry benchmark. Accuracy is expressed as the percentage of questions correct in the test for a given combination. Models were then assessed on their ability to detect hallucinations via the P(True) metric on **c)** a training set, and **d)** a hold out test set. All multiple choice questions can be found in **Dataset S3**.

**Supporting Information**

Large Language Models for Education: ChemTAsk – An Open-Source Paradigm for Automated Q&A in the Graduate Classroom

Ryann M. Perez,[a]* Marie Shimogawa[a], Yanan Chang[a], Hoang Anh T. Phan[a], Jason G. Marmorstein[a], Evan S. K. Yanagawa[a], and E. James Petersson[a]*

* Indicates Corresponding Author
[a] Department of Chemistry, University of Pennsylvania, Philadelphia, Pennsylvania 19104, USA

Ryann Perez, E. James Petersson
Email: ryper@sas.upenn.edu, ejpetersson@sas.upenn.edu

**Supporting Information Text**

**Overview**

The supporting information contains additional analysis of student surveys, usage of ChemTAsk, and self-assessment metrics as described by Farquhar *et al.* (1). Raw datasets (**Dataset 1** and **Dataset 2**) are available in comma separated values format (CSV). Code to generate the figures found in the supporting information can be found at our GitHub: https://github.com/ejp-lab/EJPLab_Computational_Projects/tree/master/ChemTAsk

**Fig. S1.** Word Cloud and top 10 most frequently used words for the body of emails submitted to ChemTAsk. Larger words in the word cloud are more frequently used. 'M' comes from acronyms such as LC/MS or MAGE.

S3

**Fig. S2.** Word Cloud and top 10 most frequently used the body of emails returned by ChemTAsk. Larger words in the word cloud are more frequently used.

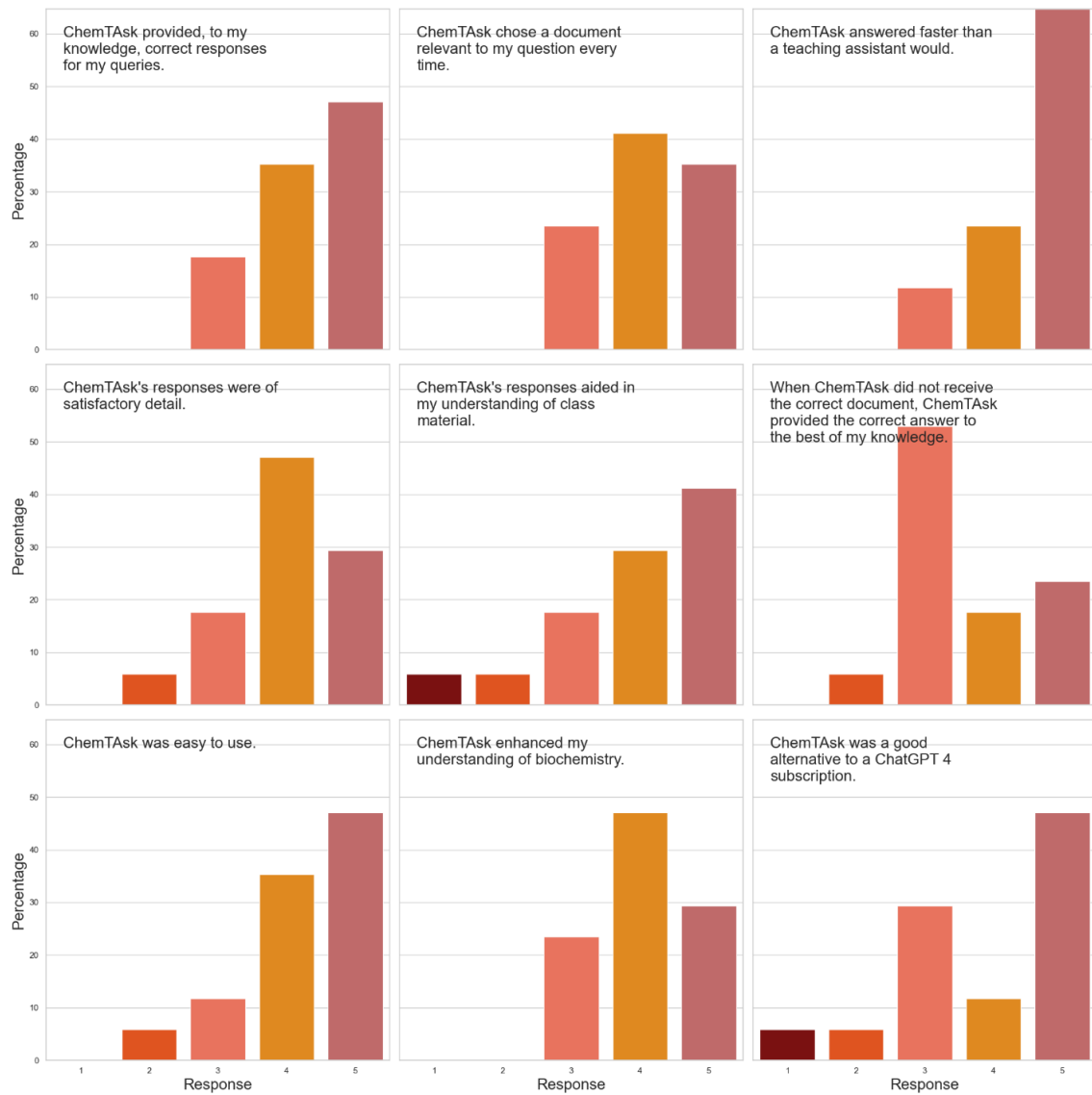

**Fig. S3.** End of study survey results (n=17). ChemTAsk was originally named Dr. ChatGPT and that name was used in the original survey.

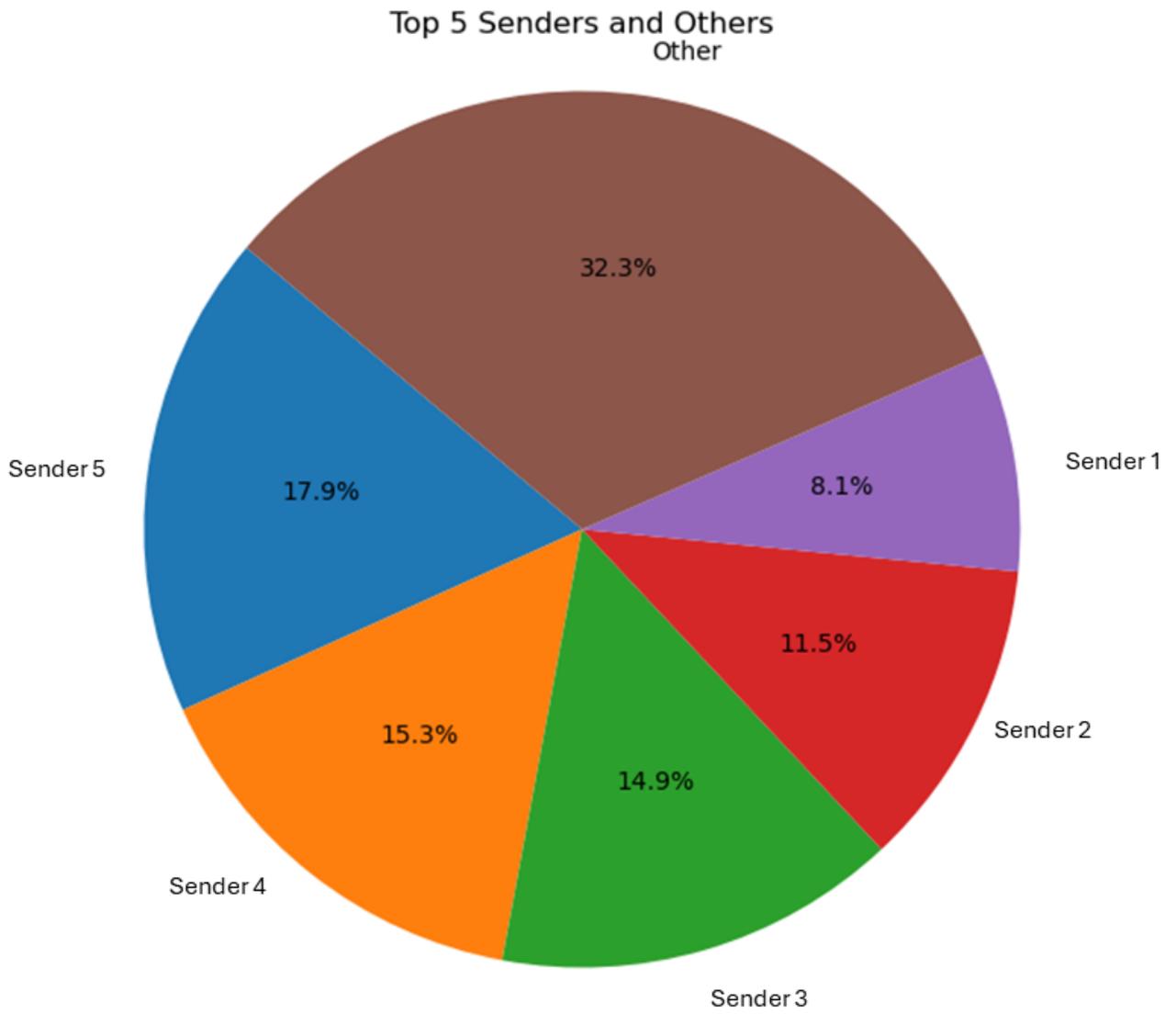

**Fig. S4.** Proportion of responses collected from the top five senders and 14 others.

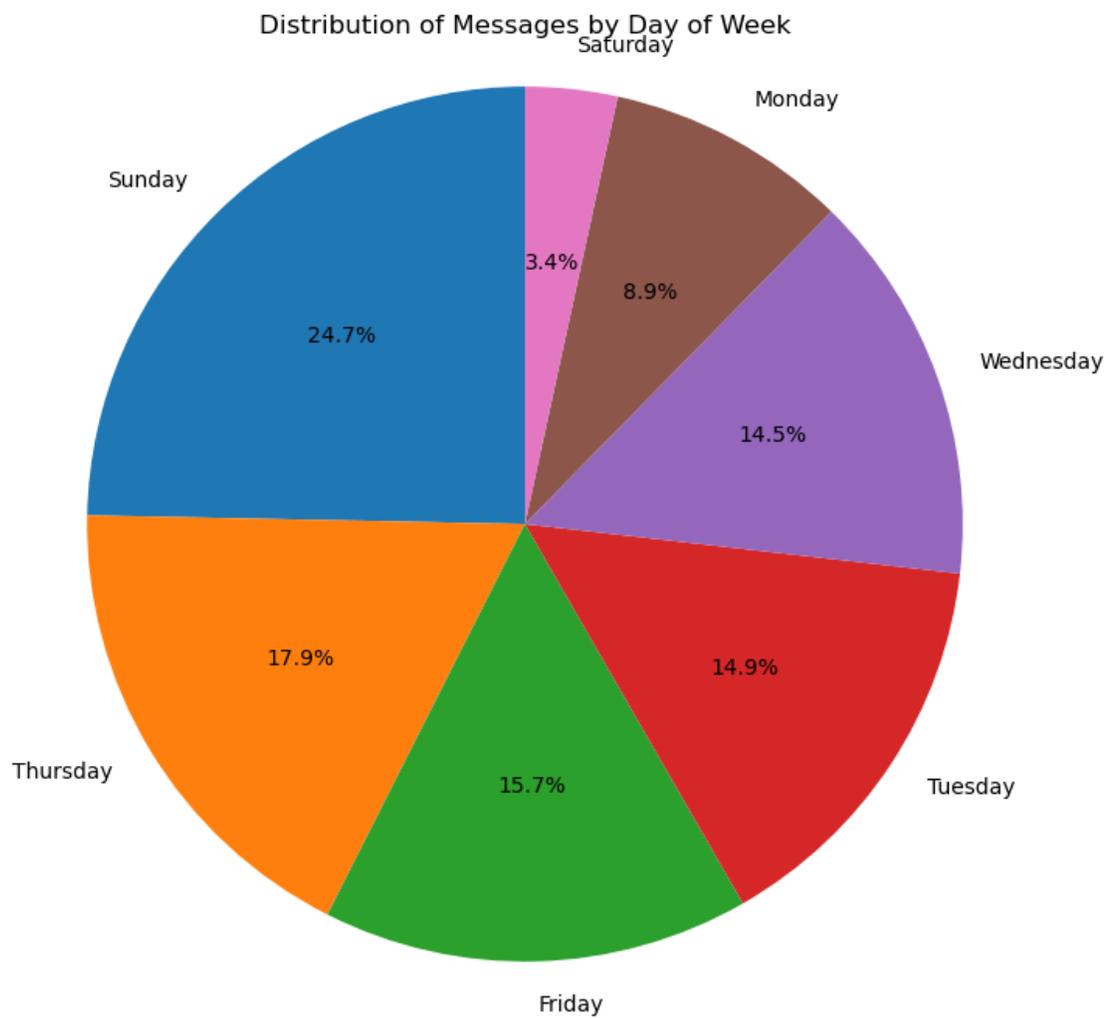

**Fig. S5.** Proportion of days ChemTAsk was utilized by students.

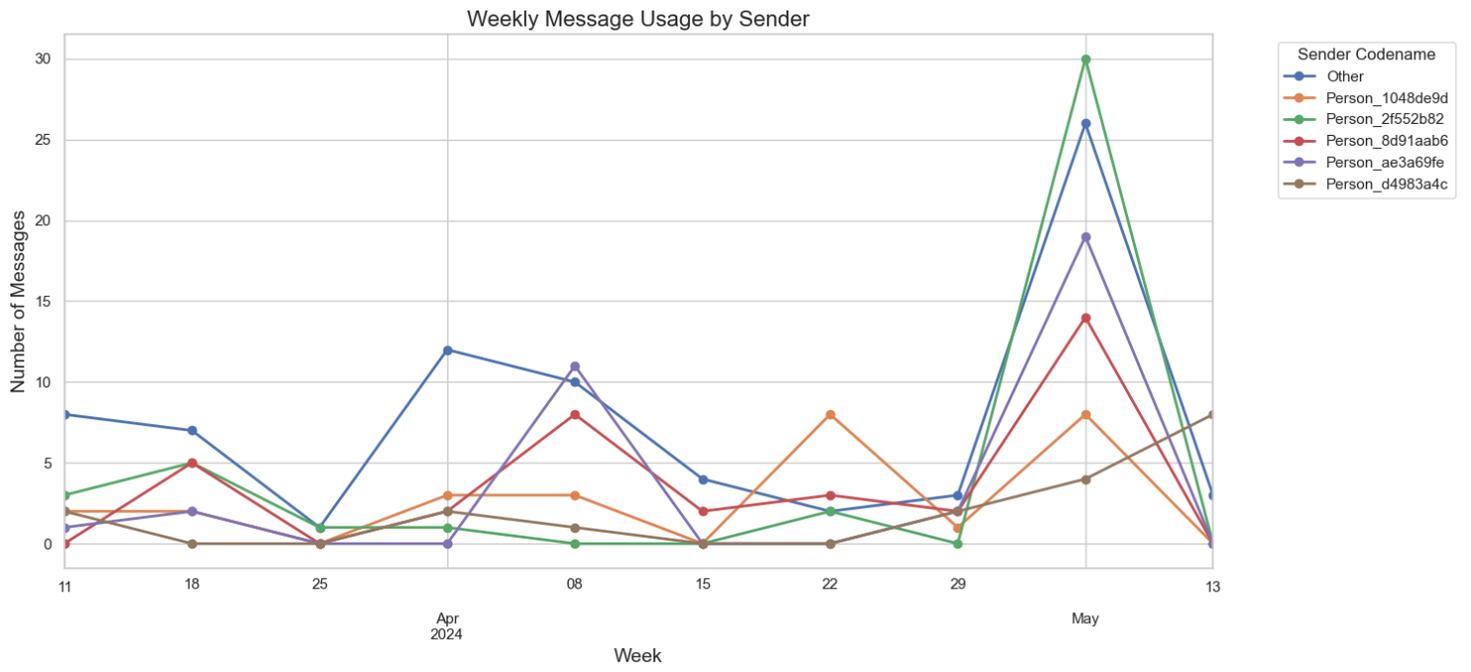

**Fig. S6.** Weekly usage over time for the top 5 senders and others.

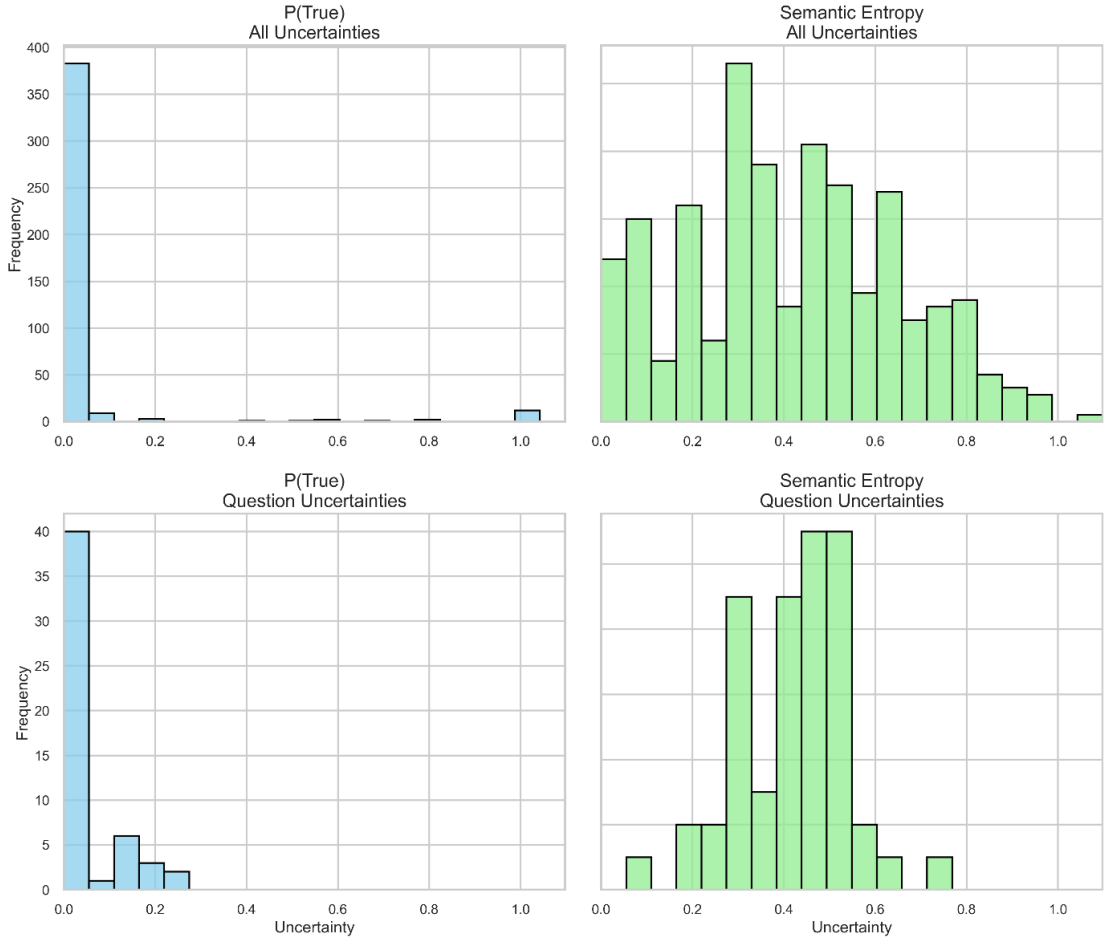

**Fig. S7.** Distribution of uncertainty measurements for paragraph level responses to student queries from ChatGPT-4o as described in (1). Top left: Distribution of uncertainty measurements from P(True) for all factual claims. Top right: Distribution of uncertainty measurements from semantic entropy for all factual claims. Bottom left: Distribution of average uncertainty on the question level from P(True). Bottom right: Distribution of average uncertainty on the question level from semantic entropy.

**SI References**